\documentclass[prb,aps,twocolumn,reprint,showpacs,preprintnumbers,amsmath,amssymb,sort&compress,superscriptaddress,floatfix]{revtex4}
\usepackage{amsbsy,bm}
\usepackage[normalem]{ulem}
\renewcommand{\Re}{\mathop{\mathrm{Re}}}
\renewcommand{\Im}{\mathop{\mathrm{Im}}}

\usepackage{graphicx}% Include figure files
\usepackage{dcolumn}% Align table columns on decimal point
\usepackage{epsfig,epstopdf}
\usepackage{subfigure}
\usepackage{color}

\begin{document}

%\preprint{a}

\title{Effects of external driving on the coherence time of a Josephson junction qubit in a bath of two level 
fluctuators}

\author{H{\aa}kon Brox}%
\affiliation{Department of Physics, University of Oslo, PO Box 1048 Blindern, 0316 Oslo, Norway}

\author{Joakim Bergli}%
\affiliation{Department of Physics, University of Oslo, PO Box 1048 Blindern, 0316 Oslo, Norway}

\author{Yuri M. Galperin}%
\affiliation{Department of Physics, University of Oslo, PO Box 1048 Blindern, 0316 Oslo, Norway}
\affiliation{Centre for Advanced Study, Drammensveien 78, Oslo, Norway 0271, Oslo, Norway}
\affiliation{A. F. Ioffe Physico-Technical Institute of Russian Academy of Sciences, 194021 St. Petersburg, Russia}

\date{\today}

\begin{abstract}
 We study the effect of external driving on the two level systems (TLSs) assumed to be a major obstacle in increasing the coherence time of solid state Josephson-junction qubits. We find, by use of a Bloch-Redfield approach, that external driving has two major effects on the TLS. The first is increased fluctuations between the two states of the TLS, the significance of this effect compared to thermal fluctuations depend on the energy splitting of the TLS compared to temperature. The second effect is a reduction in the intensity of the noise spectrum at low frequencies, and at the same time an increase in intensity around the renormalized Rabi frequency of the TLS, the driving frequency and at beatings between these two frequencies. Finally we study the ensemble averaged noise spectrum for a typical distribution of TLSs known to give origin to $\propto 1/f$ noise. We find that strong driving leads to reduced noise at low frequencies, and therefore to an increased dephasing time $T_2^Q$ of the qubit. However this effect is exponentially suppressed when the driving frequency is large compared to temperature, as we typically find for Josephson qubits. We suggest that external driving at frequencies much lower than the qubit frequency might be used in order to enhance the the qubit coherence time.
\end{abstract}

\maketitle

%%% New commands %%%
\newcommand{\eq}{\! = \!}
\newcommand{\keq}{\!\! = \!\!}
\newcommand{\kadd}{\! + \!}

\section{Introduction}
The most fundamental problem that has to be overcome in order to produce a quantum computer is the isolation of its basic elements, the quantum bits (qubits), from its environment. 
Entanglement with uncontrollable degrees of freedom is responsible for the decay of coherent superpositions of qubit states. The result is irreversible loss of the quantum information required for operation of the device.
Superconducting qubits based on the Josephson junction are leading candidates in the design of a quantum computer. They have low losses, are easily controllable by microwave pulses and can be fabricated by use of established integrated circuit technology. Recent progress in extending the decoherence time of the qubits has been achieved by identification of the sources of noise and their respective natures. This knowledge has lead to the development of countermeasures such as better isolation, as well as protocols to minimize the negative impact of the noise, see, e.g., Refs.~\onlinecite{devoret02,rebentrost,kim}, or Ref.~\onlinecite{PhysRevB.72.134519} for a review.

Bistable two level systems (TLS) existing in the tunneling junction and in the amorphous substrate used to fabricate the qubit, are thought to be the most important source of decoherence in Josephson junction qubits, \cite{martinis04,martinis05,martinis10,tian,shnirman05,zgalperinmeso,zgalperinprl} see also Ref.~\onlinecite{zzbergli}  for a review. These TLSs are assumed to give rise to the observed $\propto 1/f$ noise spectra in Josephson qubits. It is known that control and manipulation of Josephson qubits by use of microwave pulses unavoidably leads to driving of TLSs in the vicinity of the qubit. While different theories of $\propto 1/f$ noise, and their consequences have been studied in great detail, the effect of driving has with one exception been neglected.

Recently, the influence of external driving on the noise spectra of such TLS was investigated in Ref.~\onlinecite{constantin}. It was found that the noise at low frequencies was unchanged by driving, while the noise at high frequencies was  weakly reduced. 
In this article we calculate the noise spectra from a single TLS and an ensemble of driven TLSs. The results we obtain differ qualitatively from those obtained in Ref.~\onlinecite{constantin}.
The picture we arrive at is the following: In a general environment, e.g., a disordered substrate, there will be TLSs with a wide distribution of energy splittings $E$ and relaxation rates $\gamma$. Given a driving field of frequency $\Omega$, we can divide the fluctuators into two groups. Those who are far from resonance and very weakly perturbed by the driving field (group $\MakeUppercase{\romannumeral 1}$), and those who are close to resonance with the driving field (group $\MakeUppercase{\romannumeral 2}$).
We find that the TLSs belonging to group $\MakeUppercase{\romannumeral 2}$, are strongly affected by the driving provided that the driving amplitude is large compared to the relaxation rate of the TLSs. The response to the driving can roughly be described by two effects. The first is saturation of the fluctuators. A two-level fluctuator with large energy splitting compared to temperature, $E\gg k_BT$, will in the absence of driving be frozen in the ground state, with a very small probability of switching to the exited state. By driving this fluctuator with a frequency close to resonance, the probability of exitation will increase and by increasing the driving intensity the probability for the fluctuator to be found in the upper state versus the lower state will eventually be similar, thus the TLS is saturated. A driven fluctuator will thus fluctuate (much) more rapidly between its upper and lower state. Therefore the noise from this fluctuator will increase.
The second effect caused by driving is a reduction of the noise spectrum,  
\begin{align}
S_0(\omega)\propto\frac{\gamma}{\gamma^2+\omega^2}\frac{1}{\cosh^2(E/2k_BT)} \, ,
\end{align}
at frequencies centered around $\omega=0$ and at the same time increased noise at higher frequencies. The driving results in several new peaks in the noise spectrum. Most pronounced are the peaks centered at the renormalized Rabi frequency, $A'$, at the driving frequency, $\Omega$, and at beatings between these two frequencies.
We find that the net effect of strong driving is a suppressed noise spectrum at low frequencies.

A typical substrate used for fabrication of qubits is often assumed to contain TLSs with a $\propto 1/\gamma$ distribution of relaxation rates, and a smooth distribution of energy splittings, which can be approximated as uniform.~\cite{martinis05,paladino02,zzbergli} This distribution is known to give rise to $\propto 1/f$ noise at low frequencies.~\cite{kogan}
For such an ensemble of fluctuators, the low frequency noise is strongly dominated by the fluctuators with small relaxation rates. Driving at high frequencies resonant with the energy splitting of the qubit, $\hbar\Omega\approx E_Q$  where $E_Q>k_BT$ (the energy splitting needs to be large compared to temperature, in order to avoid thermal transitions between its eigenstates), will result in a significant response only from the fluctuators near resonance with the driving field. In the absence of driving, this subset of fluctuators (group $\MakeUppercase{\romannumeral 2}$) are frozen out and contribute only marginally to the ensemble averaged noise spectra, which is dominated by the fluctuators with small energy splittings, $E\leq k_BT$. Thus suppression of the low frequency noise from group $\MakeUppercase{\romannumeral 2}$ by strong driving only weakly influences the full ensemble-produced noise spectra at low frequencies, but strongly increases the noise at higher frequencies. 

However, our results show that while external driving at qubit frequency for typical ensembles of fluctuators will not have significant impact on the low frequency noise, external low frequency driving might significantly suppress it. Driving at low frequencies will effect the fluctuators that contribute most strongly to the dephasing-producing noise felt by the qubit (i.e., those with small energy splitting, $E$). These TLSs are only weakly influenced by the saturation effect since their ratio $E/k_BT$ is low and correspondingly the population level in the upper state is already high in the absence of driving. Therefore, the net effect of driving on the low $E$ fluctuators is almost entirely a shift in the frequency spectra from low to high frequencies. To us this seems like a promising method to reduce pure dephasing noise and thereby increase $T_2^Q$ for the qubit. It is, however, important to note that the high frequency noise will be increased, specifically around the Rabi frequency of the driven fluctuators as well as around the driving frequency. One should therefore make sure that these frequencies lie sufficiently far from the eigenfrequency of the qubit in order to avoid decreasing $T_1^Q$. In this article we will focus on the low frequency noise, the noise at frequencies close to the qubit splitting need to be treated separately, see Refs. \onlinecite{ustinov,martinisnature,zgalperinrabi}.

The rest of this article is divided in the following sections. In Section \ref{model} we will describe our model of a TLS in an external field and the assumptions behind it.
Thereafter in Section \ref{single} we will derive an expression for the noise spectrum from a single TLS and look at different limiting cases. In Section \ref{many} we will derive an expression for the ensemble averaged noise in the case of strong driving, for a particular distribution of TLS parameters $P(E,\gamma)\propto 1/\gamma$.
In Section \ref{sec:qubit} we will discuss the effect of the driven TLSs on the central qubit.
Finally the results will be discussed in Section \ref{discussion}.
      
\section{Model}
\label{model}

In this section we will study the dynamics of TLSs (fluctuators) subject to an external AC electric field, $\mathbf{E}_{ac}$, and a thermal environment.
The nature of the two level systems we are interested in can, e.g., be considered to be bistable fluctuators tunneling between distinct charge configurations, leading to charge noise in the qubit. These charge fluctuators might be attributed to tunneling of charges between either localized impurity states, between localized impurity states and metallic electrodes, or between different charge configurations in a dielectric material.~\cite{zgalperinmeso, phillips} We model the charge configurations associated with each state of a given TLS by its effective dipole moment $\mathbf{p}$. In order to capture the action of the environment (e.g., thermal phonons) responsible for relaxation and decoherence of the TLS, we apply the Bloch-Redfield approach.~\cite{slichter,redfield,wangsness} We assume that the interaction between different TLSs is weak compared to the coupling to the thermal bath, such that eventual correlations between the TLSs are neglected. Furthermore we assume that the TLSs couple sufficiently weakly to the qubit compared to other degrees of freedom in the environment that neglection of the qubit is justified when studying the dynamics of the TLS. This allows us to use a perturbative approach when treating the effect of the TLS(s) on the qubit. 

\subsection{Hamiltonian}
Our Hamiltonian for the TLSs closely follows that of Ref.~\onlinecite{constantin}. A fluctuator, e.g., a particle in a double well potential with associated dipole moment $\mathbf{p}$, can be modeled as a two level system with tunneling matrix element $\Delta_0$ and asymmetry energy $\Delta$.
The Hamiltonian of this TLS in an applied electric field, $\mathbf{E}_{ac}$ is then
$\bar{H}(t)=\bar{H}_0+\bar{H}_1(t)$, where $\bar{H}_0=\frac{1}{2}\left(\Delta\tau_z+\Delta_0\tau_x\right)$ and $\bar{H}_1(t)=-\tau_z\mathbf{p} \mathbf{E}_{ac}(t)$. Here $\tau_{x,z}$ are the Pauli matrices and $\mathbf{E}_{ac}(t)=\mathbf{E}_{ac}\cos{\Omega t}$ is an AC electric field of angular frequency $\Omega$ coupling to the electric dipole moment of the TLS. Furthermore, the TLS interacts with the qubit through $\bar{H}_{F-Q}$ and couples to the environment through $\bar{H}_{F-env}$.

By diagonalization of $\bar{H}_0$, the Hamiltonian in the energy eigenbasis becomes
\begin{align}
H&=H_0+H_1(t)+H_{F-Q}+H_{F-env},\nonumber\\
H_0&=\frac{1}{2}E\sigma_z,\nonumber\\
H_1(t)&=-\eta(\Delta\sigma_z+\Delta_0\sigma_x)\cos{\Omega t},\nonumber\\
H_{F-Q}&=v \mu_z\otimes \tau_z=v \mu_z\otimes\left( \frac{\Delta}{E}\sigma_z+\frac{\Delta_0}{E}\sigma_x\right)
\label{Ham}
\end{align}
where $E=\sqrt{\Delta^2+\Delta_0^2}$ and $\eta=\mathbf{p}\mathbf{E}_{ac}/E$. The matrices $\mu_z$ and $\sigma_{x,z}$ are Pauli matrices acting in the eigenbasis of the qubit and the TLS, respectively and $v$ is the qubit-fluctuator coupling parameter.
The TLS-qubit coupling will be neglected when treating the dynamics of the TLS, assuming it is weak compared to other terms. But it is, of course, important with regards to the decoherence of the qubit. We note that the situation when the qubit and the TLS have very close splittings is an exception. Then the interaction is strong. See, e.g., Refs.~\onlinecite{ustinov,martinisnature,zgalperinrabi}. 

Rather than specifying the explicit nature of the coupling to the environment, $H_{F-env}$, we make use of the Bloch-Redfield equation, where the environment enter as damping terms, seeking to relax the density matrix towards its thermal equilibrium value.
The Bloch-Redfield equations for the density matrix elements of the two level system in the eignenbasis of $H_0$ are:~\cite{slichter}
\begin{equation}
\dot{\rho}_{\alpha\alpha'}=%\nonumber\\
%& \ \ 
\frac{i}{\hbar}\langle\alpha|[\rho,H]|\alpha'\rangle+\sum\limits_{\beta,\beta'}R_{\alpha\alpha',\beta\beta'}\left(\rho_{\beta \beta'}-\rho_{\beta \beta'}^{eq}\right).
\label{BR}
\end{equation}
The rates $R_{--,++}=R_{++,--}\equiv T_1^{-1}$ and $R_{-+,-+}=R_{+-,+-} \equiv T_2^{-1}$ can be derived from perturbation theory,~\cite{slichter} and the equilibrium density matrix, $\rho_{\beta,\beta'}^{eq}(T)$ is introduced phenomenologically in order to achieve relaxation towards thermal equilibrium.

By use of the rotating wave approximation, we can simplify the first two terms of our Hamiltonian, Eq. \eqref{Ham}, obtaining
\begin{equation}
H_{RWA}=\frac{1}{2}E\sigma_z-\frac{\eta\Delta_0}{2}\left(e^{-i\Omega t}|+\rangle\langle -|
+e^{i\Omega t}|-\rangle\langle +|\right).
\label{hrwa}
\end{equation}
Inserted into the Bloch-Redfield equation Eq.\eqref{BR} we find that the time evolution of the elements of the density matrix is governed by the following set of differential equations:
\begin{align}
\frac{d\rho_{++}}{dt}&=\frac{i\eta\Delta_0}{2\hbar}\left(e^{-i\Omega t}\rho_{-+}-e^{i\Omega t}\rho_{+-}\right)-\frac{1}{T_1}\left(\rho_{++}-\rho_{++}^{eq}\right)\nonumber\\
 \frac{d\rho_{-+}}{dt}&=\frac{iE}{\hbar}\rho_{-+}+\frac{i\eta\Delta_0}{2\hbar}e^{i\Omega t}(2\rho_{++}-1)
-\frac{1}{T_2}\rho_{-+} \, .
\label{diffeq}
\end{align}
Here we note that $\rho_{+-}=\rho_{-+}^{\dagger}$ and $\rho_{--}=1-\rho_{++}$.
To avoid the explicit time dependence we make the transformation 
$f=e^{-i\Omega t}\rho_{-+}$ and $f^*=e^{i\Omega t}\rho_{+-}$.
 We also introduce the Rabi frequency $A=\eta\Delta_0/\hbar$ and the deviation from resonance $z=E/\hbar-\Omega$.
Furthermore we make the approximation for the relaxation rates $\gamma=1/T_1=1/T_2$. While not valid in general, this approximation is believed to be valid when the decoherence is isotropic.~\cite{wang} The general relationship  $T_2\leq 2T_1$ can be derived from the master equation approach.~\cite{breuer,slichter} Thus by making this simplifying assumption, asymmetry of the relaxation behavior of the TLS are left out. However we believe that these details are not of crucial importance for the results derived concerning the qubit's decoherence due to the TLSs.

\section{Single TLS}
\label{single}
In this section we will first solve the equations of motion for a single TLS and then proceed to find its noise spectrum. This we will analyze later when we study the influence of the TLS(s) on the qubit. Using the notations $N=\rho_{++}$, $\Re{
f}=R$ and $\Im{f}=I$ 
one can cast the Bloch-Redfield equation, Eq.~\eqref{diffeq}, in the form
\begin{align}
\dot{N}&=-AI-\gamma \left(N-N_{\text{eq}}\right) \, ,\nonumber\\
\dot{I}&=A(N-1/2)-\gamma I+z R \, ,\nonumber\\
\dot{R}&=-z I-\gamma R \, .
\label{diffeq2}
\end{align}
Solution of  Eqs. \eqref{diffeq2} can be written as
\begin{align} \label{ns}
 \left( \begin{array}{c}
N(t)\\
R(t)\\
I(t)\end{array} \right) 
= \Lambda(t)\left( \begin{array}{c}
N_0 \\
R_0 \\
I_0 \end{array} \right) +\bm{\kappa}
\end{align} 
where $N_0$, $R_0$ and $I_0$ are the initial values of $N$, $R$ and $I$, respectively. 
The solution of the homogeneous part of the equation is given by
\begin{equation}
\Lambda
= e^{-\gamma t} \!\! \left( \begin{array}{ccc}
\frac{A^2\cos{A't}+z^2}{A'^2}&\frac{zA\left(\cos{A't}-1\right)}{A'^2}&-\frac{A\sin{A't}}{A'}\\
\frac{zA\left(\cos{A't}-1\right)}{A'^2}&\cos{A't}&-\frac{z\sin{A't}}{A'}\\
\frac{A\sin{A't}}{A'}&\frac{z\sin{A't}}{A'}&\cos{A't} \end{array} \right),
\label{transient}
\end{equation} 
where $A'=\sqrt{A^2+z^2}$ is the renormalized Rabi frequency.
The particular solution is given by
\begin{align}
\bm{\kappa}=\left( \! \!
\begin{array}{c}
N_{\text{eq}}+\frac{A^2\left(N_{\text{eq}}-\frac{1}{2}\right)}{\gamma^2+A'^2}\left[e^{-\gamma t}\left(\cos{A't}+\frac{\gamma\sin{A't}}{A'}\right)-1\right]\\
\frac{zA\left(N_{\text{eq}}-\frac{1}{2}\right)}{\gamma^2+A'^2}\left[e^{-\gamma t}\left(\cos{A't}+\frac{\gamma\sin{A't}}{A'}\right)-1\right]\\
\frac{\gamma A\left(N_{\text{eq}}-\frac{1}{2}\right)}{\gamma^2+A'^2}\left[1-e^{-\gamma t}\left(\cos{A't}-\frac{A'\sin{A't}}{\gamma}\right)\right] \end{array}\! \! \right) \! \!.
\label{steadystate}
 \end{align}
We note that by setting $t\rightarrow \infty$ in Eq.\eqref{steadystate} it is possible to directly read out the steady state solution.

\subsection{Noise spectrum from a single TLS}

Given the above specified qubit-TLS coupling, Eq.\eqref{Ham}, the TLS is only responsible for pure dephasing of the qubit ($T_2$ processes) and cannot induce transitions between the eigenstates of the $\mu_z$ operator.
The dynamics of the TLS leads to uncontrolled fluctuations in the energy splitting of the qubit, leading to an uncertainty in its phase. 
Alternatively, it leads to entanglement both directly to the fluctuators and indirectly to the environment of the fluctuators. In this article we will analyze the effect of the fluctuators on the qubit through the two-time correlation function of the operator responsible for the noise in the qubit energy splitting.~\cite{girvinrmp}
We define it as
\begin{align}
G(t_1,t_2)&=\left\langle\left[q(t_2)-\bar{q}(t_2)\right]\left[q(t_1)-\bar{q}(t_1)\right]\right\rangle\nonumber\\
&=\sum_j\langle q(t_2)-\bar{q}(t_2)\rangle|_{q_j(t_1)-\bar{q}(t_1)}\nonumber\\
&\quad\times[q_j(t_1)-\bar{q}(t_1)]P[q_j(t_1)-\bar{q}(t_1)].
\label{corr1}
\end{align}
Here $q_j(t)$ is a realization of a measurement at time $t$ of the operator $v\tau_z$, giving the variation in the qubit's energy splitting due to its interaction with a TLS. While $\bar{q}(t)=\langle q(t)\rangle=\sum_j q_j(t)P[q_j(t)]$ is the ensemble 
average of $q(t)$, and $P[q(t)]$ is the probability distribution of $q$ at time $t$. 
The Bloch-Redfield equations, Eq.\eqref{diffeq2}, give the average time evolution of an ensemble of systems with the same initial condition, averaged over the details of the uncontrolled decoherence processes. Thus we find that $\langle q(t)\rangle|_{q_j}$, is simply the solution of the Bloch-Redfield equations, given the initial value $q_j$.
More explicitly, we find the following expression:
\begin{equation}
\langle q(t)\rangle|_{q_j}=\frac{2 v}{E}\big\{ \Delta N(t)-\Delta_0\left[R(t)\cos{\Omega t}-I(t)\sin{\Omega t}\right]  \!\big\},
\label{Poperator}
\end{equation}
where the initial condition $q_j$ is written in terms of the initial values $N_0$, $R_0$ and $I_0$.
The corresponding expression for $\bar{q}(t)$ is
\begin{equation}
\bar{q}(t)=\frac{2 v}{E}\left[\Delta N_{\text{ss}}-\Delta_0\left(R_{\text{ss}}\cos{\Omega t}-I_{\text{ss}}\sin{\Omega t}\right)\right],
\label{Poperatorbar}
\end{equation}
where $N_{\text{ss}}$, $R_{\text{ss}}$ and $I_{\text{ss}}$ are the steady state limits for $N(t)$, $R(t)$ and $I(t)$, respectively, obtained from Eq.~\eqref{ns} at $t \to \infty$.

Thus we find that the correlator defined by Eq.~\eqref{corr1} depends on the phase of the driving field at both $t_1$ and $t_2$. In a qubit experiment, where $t_1$ and $t_2$ are the initialization and measurement time, respectively, we assume that we do not have sufficient control over the phase of the driving field at initialization time and therefore average over the initial phase of the driving field. The details of this procedure are given in Appendix \ref{field}.

\begin{figure}[htb]
  \includegraphics[width=0.35\textwidth]{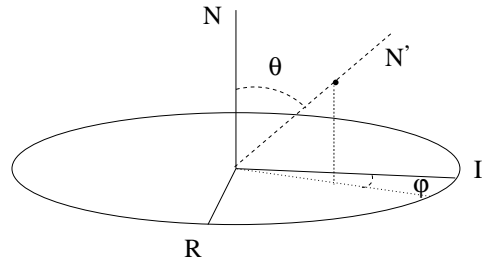}
 \caption{The coordinate transform used in order to diagonalize the density matrix.
Here $N$, $R$ and $I$ are the parameters determining the density matrix in the energy eigenbasis of the TLS, while $N'$, $R'$ and $I'$ denote the same parameters in the rotated frame defined by Eqs.\eqref{ctrans}. The frame is defined such that the off diagonal elements of the density matrix vanish in steady state, i.e., $R'_{\text{ss}}=I'_{\text{ss}}=0$. In external driving, the steady state values of the off-diagonal elements of the density matrix are in general non-zero. The transform is used in order to make use of the average procedure described in Appendix \ref{App:init}. In the absence of driving the two frames coincide.}
\label{cootrans}
\end{figure}

The details of the procedure used to calculate the two time correlation function, Eq.~\eqref{corr1}, by use of the Bloch-Redfield formalism is described in detail in Appendix \ref{App:init}. The full procedure includes a coordinate transform, and in the following we give an outline of the procedure.
First we find the density matrix in the steady state that we might visualize as a point within the Bloch sphere.
Next, we note that in an external field the steady state solution can in general lie anywhere in the Bloch sphere and not necessarily along the z-axis. Therefore, it is necessary to transform to the coordinate system where the steady state solution lies along the z-axis. The angles defining this transform are given by the steady state solution of Eqs.(6-8), and are illustrated in Fig.~\ref{cootrans}. The angles $\theta$ and $\phi$ are defined by the relations:
\begin{equation} \label{ctrans}
\tan \theta=\frac{\sqrt{R_{\text{ss}}^2+I_{\text{ss}}^2}}{1-2N_{\text{ss}}}, \quad \tan \phi= \frac{R_{\text{ss}}}{I_{\text{ss}}}
\end{equation}
The details of this transform and its application to the evaluation of the two time correlation function is described in Appendix \ref{App:init}.
Since the off-diagonal elements of the density matrix vanish in this choice of basis, we are allowed to use the states $|-'\rangle$ and $|+'\rangle$ in the rotated basis as initial states, weighted by the mean population levels obtained from the density matrix in the steady state, that gives us $P[q]$.

In the absence of external driving the density matrix in the equilibrium will always lie along the z-axis and the ensemble average of $R$ and $I$ vanishes. Thus we do not require the coordinate transform. In this particular case, after introducing $\tau=t_2-t_1$, the explicit expression for the two-time correlation function given by Eq.~\eqref{corr1} is
\begin{align}
G(t_1,t_2)&=\frac{4\Delta^2v^2}{E^2}\left\langle\left[N(t_2)-N_{\text{eq}}\right]\left[N(t_1)-N_{\text{eq}}\right]\right\rangle\nonumber\\
&=\frac{4\Delta^2v^2}{E^2}\lambda_{11}(\tau)N_{\text{eq}}\left(1-N_{\text{eq}}\right).
\label{g}
\end{align}
Here $\lambda_{11}(\tau)$ denotes the $11$ element of $\Lambda(\tau)$  given by Eq.~\eqref{transient}. The dependence of $N(t_2)$ on the initial values $R(t_1)$ and $I(t_1)$ vanish in the absence of external driving. We can therefore in this simple case write the propagator $\Lambda(\tau)$ as a scalar function $\lambda_{11}(\tau)$. 

In the general case, when driving is included, we find
\begin{equation} \label{pop}
G(t_1,t_2) \propto f(\tau,A,\gamma,z) N'_{\text{ss}}(1-N'_{\text{ss}}) . 
\end{equation}
Here  $f(\tau,A,\gamma,z)$ describes the dynamics of the density matrix, while
$N'_{\text{ss}}$ is the population of the upper level in the rotated frame, illustrated in Fig.1. We have
\begin{equation} \label{pop1}
N'_{\text{ss}}(1-N'_{\text{ss}})=N_{\text{ss}}(1-N_{\text{ss}})
+g(A,\gamma,z)\left(N_{\text{ss}}-1/2\right)^2 
\end{equation}
where $g(A,\gamma,z)$ describes the details of steady state density matrix ($g(A,\gamma,z)=0$ if $R_{\text{ss}}=I_{\text{ss}}=0$).
See Eq.~\eqref{corrfinal} and Eq.~\eqref{nss} for details.

From the correlation function, Eq.~\eqref{corr1}, we can compute the contribution of a single TLS to the noise spectrum acting on the qubit.
The spectrum is given by the expression:
\begin{align}
S(\omega)&=\sqrt{\frac{2}{\pi}}\int\limits_{-\infty}^{\infty}e^{i\omega\tau}G(|\tau|,0)d\tau,
\end{align}
where we took into account that the correlation function $G(t_1,t_2)$ is translation invariant after the averaging procedure described in Appendix \ref{App:init}. We note that the irreversible Bloch-Redfield equations require the measurement time to succeed the preparation time, therefore we need the absolute value of $|\tau|$ in the definition.
From the full spectrum at arbitrary frequency, given by Eq.~\eqref{corr6} in Appendix \ref{fullspectrum}, we obtain in the limit $\Omega>\gamma$ the following expression for $S(\omega)$: 
\begin{eqnarray} \label{smallw}
&&S(\omega)=8\sqrt{\frac{2}{\pi}}\left(\frac{v\Delta}{EA'}\right)^2N'_{\text{ss}}(1-N'_{\text{ss}}) \left\{\! \! \phantom{\frac{}{}}a_1L(\omega) \right. \nonumber \\ 
&& \  \left.
+\sum\limits_{\pm}\left[a_2L(\omega\pm A')\pm a_3\frac{\omega\pm A'}{\gamma}L(\omega\pm A')\right]  \!\right \}
\end{eqnarray}
where $L(\omega)=\gamma/(\gamma^2+\omega^2)$, 
\begin{eqnarray*}
a_1&=&z^2\cos^2\theta-zA\sin\theta\cos\theta\cos\phi, \\
a_2&=&A^2\cos^2\theta+zA\sin\theta\cos\theta\sin\phi, \\
a_3&=&AA'\sin\theta\cos\theta\cos\phi .
\end{eqnarray*}
In this limit, the spectrum only contains peaks at zero frequency $\omega=0$ and at the Rabi frequency $A'$.

\subsection{Low frequency noise}

In the remaining part of this article we are interested in the noise at low frequency, and how it is changed by 
the external driving. The reason behind this focus is that the noise at low frequencies has been identified as the dominant source of pure dephasing in Josephson junction qubits. It is known that for the
Gaussian noise and diagonal qubit fluctuator coupling, the off-diagonal elements of the qubit density matrix relaxes 
(in the case of the free induction decay) at a rate~(see, e.g., Ref.~\onlinecite{zzbergli}) 
\begin{align}
\frac{1}{T_2^Q}\propto\int\limits_{-\infty}^{\infty}\frac{\sin^2(\omega t/2)}{\omega^2}S(\omega) \, d\omega .
\label{T2}
\end{align}
For long measurement times this distribution becomes narrow, such that the pure dephasing
rate of the qubit is given by $1/T_2^Q=\pi S(0)$ when $t\rightarrow\infty$. However, for realistic measurements with finite measurement times, the dephasing rate is determined by the noise spectrum 
in a finite domain of low frequencies centered at $\omega=0$. 
In the following, we will first derive expressions for the 
low-frequency contribution to  $S(\omega)$ due to a single TLS in the absence of external driving ($A \to 0$) and in the case of strong driving ($A \gg \gamma$).
These expressions will later be used to derive the low frequency noise spectrum for a specific distribution of driven TLSs.

From Eq.~\eqref{smallw} we find a crossover from a regime where the driving contributes as a weak perturbative effect to a regime strongly dependent on the driving, the crossover takes place around $A\approx |z|$, given that $A \gg \gamma$.
We proceed by deriving the limiting expressions in the resonant region $|z|\ll A$ and in the off resonant region $|z|\gg A$. 
Using the expression $N_{\text{eq}}=\left(e^{E/kT}+1\right)^{-1}$ and Eq.~\eqref{smallw}
we get in the off resonant regime, $|z|\gg A$,
\begin{align} 
S_{A\gg\gamma}^{(or)}(\omega)&\approx\sqrt{\frac{8}{\pi}}\left(\frac{v\Delta}{E}\right)^2\frac{L(\omega)}{\cosh^2(E/2kT)}\nonumber\\
&\quad\times\left[\left(1-\frac{5A^2}{4z^2}\right)+\frac{7A^2}{4z^2}\sinh^2{\frac{E}{2kT}}\right].
\label{zexpy}
\end{align}
As follows from the above expression, the driving only weakly [$\propto (A/z)^2 \ll 1$] affects the fluctuators that are far from resonance.

The corresponding leading contribution to the noise spectrum in the resonant regime $|z|<A$, is
\begin{align} 
S_{A\gg\gamma}^{(res)}(\omega)&\approx 5\sqrt{\frac{2}{\pi}} \left(\frac{v\Delta}{E}\right)^2\frac{\gamma}{A^2}.
\label{strong2}
\end{align}
From the full spectrum Eq.~\eqref{corr6}, together with Eq.~\eqref{pop1} for the population of the density matrix, we can identify two main effects of external driving on the TLS noise spectra. The first effect is the altered equilibrium population of the density matrix due to driving. We can, by use of Eq.~\eqref{steadystate}, express the occupation of the density matrix in steady state by 
\begin{equation}
N_{\text{ss}}=\frac{1}{2}+\frac{N_{\text{eq}}-1/2}{1+(A^2/\gamma)L(z)},
\end{equation}
 as previously found in Ref.~\onlinecite{constantin}.
\begin{figure}[htb]
  \includegraphics[width=\columnwidth]{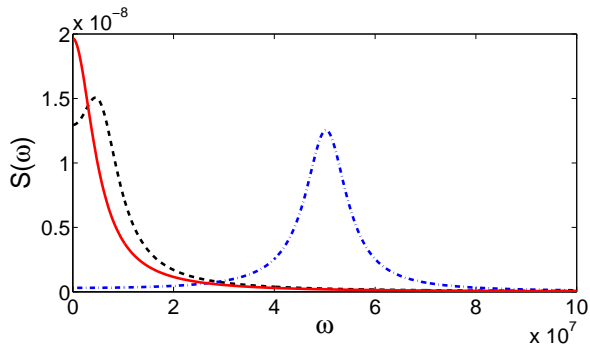}
 \caption{Noise spectrum induced by a single TLS 
for $A=0$ (solid line), $5\times 10^6$ Hz (dashed line) and  $5\times 10^7$ Hz (dash-dotted line).
We see that with this choise of parameters, driving reduces the noise at zero frequency, but enhances the noise at higher frequencies. The peak at the renormalized Rabi frequency, $A'$, is the most pronounced. Peaks at higher frequencies are suppressed as long as $A\ll\Omega$. The parameters used in the figure are $T=0.2$ K, $\Omega=E/\hbar=10^{10}$ Hz, $\gamma=5\times 10^6\; \text{s}^{-1}$, $\gamma_0=10^7\; \text{s}^{-1}$, }
  \label{shiftspec}
\end{figure}
\begin{figure}[htb]
  \includegraphics[width=\columnwidth]{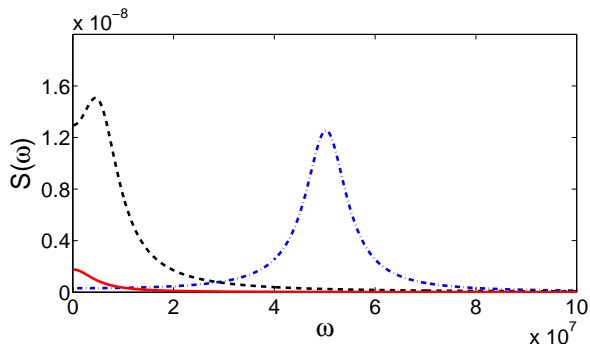}
 \caption{Noise spectrum induced by a single TLS, with the same parameters as used in Fig.\ref{shiftspec}, but at lower temperature $T=0.05$ K. At this temperature, the noise is weak in the absence of the driving (solid line) since the fluctuator is frozen in its ground state, $N_{\text{eq}}\approx 0.01$. Here the main effect of driving with strenght $A=5\times 10^6$ Hz (dashed) is to increase the probabiliy of excitation leading to increased noise at all frequencies. When the strength of the driving is $A=5\times 10^7$ Hz (dashed/dots) the noise spectrum is shifted sufficiently away from $\omega=0$ towards $\omega=A'$ such that the effect of increased fluctuations is offset by the shift in the spectrum. Thus the noise at $\omega=0$ is reduced.}
  \label{shiftspecT}
\end{figure}

External driving results in saturation of the steady state density matrix when $A\approx\gamma$ for $|z|\ll \gamma$ and when $A\approx |z|$ for $|z|\gg \gamma$. This saturation contributes to increased fluctuation rate between the upper and lower level of the TLS, see Fig.~\ref{shiftspec}, and therefore this effect contributes to a higher intensity of the noise at all frequencies (that was not found in Ref.~\onlinecite{constantin}). This is especially true for TLSs where the energy splitting is large compared to temperature, meaning that the noise is very weak in thermal equilibrium since the system spends almost all its time in the ground state (see Fig.~\ref{shiftspecT}).
There is, however, another very pronounced effect, not caught by the model of Ref.~\onlinecite{constantin}.
The full noise spectrum of the driven TLS Eq.~\eqref{corr6} is composed of several peaks. At low frequencies the most pronounced are the one centered around $\omega=0$ and the two peaks centered around the renormalized Rabi frequency, $\omega=\pm A'$, see Fig.~\ref{shiftspec}. From the last term of Eq.~\eqref{smallw} we see that the intensity around the $\omega\approx 0$ peak is reduced when the Rabi frequency $A$ becomes comparable in magnitude to the deviation from resonance  $z$. Thus we have a shift in the intensity from low frequencies to frequencies around the Rabi frequency. This shift might be beneficial in reducing the low frequency noise responsible for pure dephasing of the qubit.~\cite{zgalperinprl} In a simplified picture the total noise at a given frequency as a response to external driving can therefore be regarded as a result of two competing mechanisms, the increased fluctuations due to increased population in the upper level of the TLS, and the shift in the spectrum from low to high frequencies. At sufficiently strong driving this shift leads to a reduction in 
the low-frequency  ($\omega \lesssim \gamma$) contribution to the noise spectrum  from a TLS with $E \lesssim kT$ as 
\begin{align}
\frac{S(\omega \lesssim \gamma)_{A\gg\gamma}^{(res)}}{S(\omega \lesssim \gamma)_0}=\frac{5\gamma^2 
}{2A^2} \, .
\end{align}

\section{Ensemble of TLS\lowercase{s}}
\label{many}

In this section we will analyze the noise from an ensemble of the fluctuators studied in the preceding section. Our purpose is to roughly estimate the effect of external driving on the noise spectrum for a realistic Josephson-junction qubit experiment.
We will more specifically assume the following distribution of 
the TLS parameters.
First, we note again that our calculations are based on the assumption that the dynamics of the fluctuators are independent of the state of the qubit. Furthermore, we assume in the following that the strength of the fluctuator-qubit coupling $v$ is uncorrelated with the relaxation rate $\gamma$ and the energy $E$. 
Assuming that $\Delta_0$ is an exponential function of an almost uniformly distributed parameter, such as tunnel barrier height,~\cite{phillipsart,halperin,zzbergli}  the distribution of 
the
TLS parameters becomes
\begin{align}
P(\Delta,\Delta_0)&=P_{\text{TLS}}/\Delta_0
\label{dist1} 
\end{align}
where $P_{\text{TLS}}$ is proportional to the density of states per unit energy and volume.
This distribution is already widely used in models of decoherence in qubits, where it is known to give origin to the $\propto 1/\omega$ dependence of the noise spectrum at low frequencies.

Since in the following it is more convenient to work with the relaxation rate $\gamma$ and the unperturbed fluctuator energy $E$, we recast Eq.~\eqref{dist1} by use of the relationship $\gamma=\gamma_0(E)(\Delta_0/E)^2$, where $\gamma_0(E)$ is the maximum relaxation rate for a fluctuator of energy $E$, obtaining
\begin{align}
P(E,\gamma)&=\frac{P_{\text{TLS}}}{\gamma\sqrt{1-\gamma/\gamma_0(E)}},
\label{dist}
\end{align}
for $\gamma\in[\gamma_{\min},\gamma_0(E)]$.
The distribution has to be cut at the relaxation rate $\gamma_{\min}$ of the slowest fluctuator. 
However, we find that the noise spectra at frequencies $\omega \gg \gamma_{\min}$, and therefore measurements carried out with the
measurement time $\tau=1/\omega \ll 1/\gamma_{\min}$, are not sensitive to the cutoff. 
The maximal relaxation rate, $\gamma_0$, is a power-law function of the energy $E$. Since in the following we restrict ourselves to order-of-magnitude estimates we will replace $\gamma_0(E)$ by a constant rate,  $\gamma_0 \approx \gamma_0(kT)$.
Our calculations (see Appendix \ref{strongdriving}) show that using this assumption the noise spectrum depends only weakly on $\gamma_0$. 

Before we proceed to evaluation of ensemble integrals it is convenient to introduce a new variable
\begin{align}
a&=A/\sqrt{\gamma}=\mathbf{E}_{ac}\mathbf{p}/\hbar\sqrt{\gamma_0}.
\end{align}
This variable is independent of $\gamma$ and can be treated like a constant when integrating over distributions of TLSs.

If we assume that the qubit-fluctuator coupling $v$ is uncorrelated with $\gamma$ and $E$, and by using that the single fluctuator spectrum is $\propto v^2$, we can express the ensemble averaged noise spectrum by
\begin{align}
\bar{S}_a(\omega)&=\langle v^2\rangle\int\limits_0^{E_{max}}\int\limits_{\gamma_{\min}}^{\gamma_{0}}S_a(\omega,E,\gamma)P(E,\gamma) \, d\gamma dE.
\label{ens}
\end{align}
Here we have introduced the notation $\langle v^2\rangle=\int\limits_{v_{\min}}^{v_{max}}v^2P(v)dv$, where $P(v)$ is the distribution of the qubit-fluctuator coupling $v$.
\begin{figure}[htb]
  \includegraphics[width=0.4\textwidth]{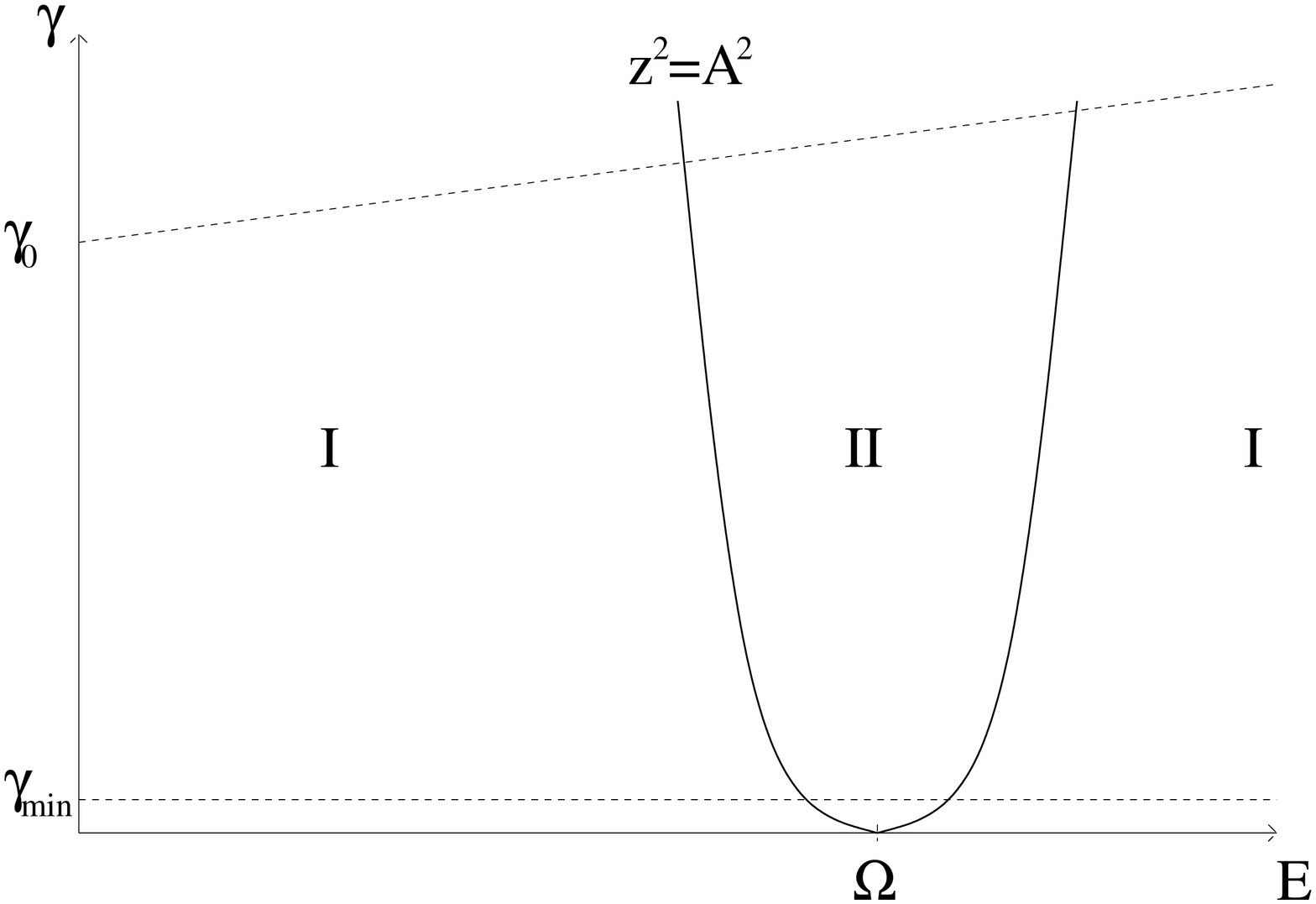}
  \caption{The full TLS parameter domain. The TLSs inside the parabola $(\Omega-E/\hbar)^2\leq A^2$, group $\MakeUppercase{\romannumeral 2}$, are resonant with the applied field. For driving frequencies $\Omega>k T/\hbar$ the major contribution to noise origin from low-$E$ fluctuators belonging to group $\MakeUppercase{\romannumeral 1}$ outside the resonant sector. The contribution to the noise due to these fluctuators are not changed significantly by driving at frequencies much higher than their energy splittings $E$.} 
\label{baseintfig}
\end{figure}
By use of the given distribution of TLSs, Eq.~\eqref{dist}, and the expression for the noise in zero driving, Eq.~\eqref{zexpy}, we can evaluate the ensemble averaged noise in the case of no driving ($a=0$). The detailed calculation is given in Appendix~\ref{strongdriving}. 
Using the relationship $\Delta^2/E^2=1-\gamma/\gamma_0$ between the relaxation rates and the fluctuator potential parameters  \cite{zzbergli} we find that the averaged spectral density is given by
\begin{equation} \label{ensfree}
\bar{S}_0(\omega) \approx \sqrt{\frac{8}{\pi}}\langle v^2 \rangle kT P_{\text{TLS}} \left\{ \begin{array}{lr}
\omega^{-1}, & \gamma_{\min}<\omega<\gamma_0, \\
\gamma_{\min}^{-1}, & \omega \lesssim \gamma_{\min}. \end{array} \right.
\end{equation}
We conclude that without driving we obtain noise $\propto 1/\omega$ for the interval $\gamma_{\min}<\omega<\gamma_0$ that turns over to a constant value for $\omega<\gamma_{\min}$. 

Next we proceed to strong driving, which we have defined by $a^2>\gamma_0$. In order to evaluate the noise in this regime we split the domain of integration in two parts, see Fig.~\ref{baseintfig}. The resonant domain  $a^2\gamma>z^2$ (group $\MakeUppercase{\romannumeral 2}$), where the fluctuators are strongly affected by the external field $\mathbf{E}_{ac}$, and the offresonant domain $a^2\gamma<z^2$ (group $\MakeUppercase{\romannumeral 1}$), where the fluctuators does only weakly respond to the driving field.
We approximate the full integral by using the asymptotic limits given by the undriven noise spectra, Eq.~\eqref{zexpy}, in the offresonant domain, while the strong driving limit is given by Eq.~\eqref{strong2}. The total ensemble averaged noise spectra for frequencies in the interval $\gamma_{\min}<\omega<\gamma_0$ is (see Appendix~\ref{strongdriving} for details of derivation)
\begin{equation} \label{ensdriven}
\bar{S}_{a^2>\gamma}(\omega)\approx \sqrt{\frac{8}{\pi}}\langle v^2\rangle P_{\text{TLS}} \! 
\left( \! \!\frac{k_BT}{\omega}-\frac{4\hbar a}{\cosh^2{\frac{\hbar\Omega}{2kT}}}\frac{1}{\sqrt{\omega}} \!\right)
\end{equation}
for $\hbar a\sqrt{\omega} \ll k T$ and $\sqrt{\gamma_0}<a<\Omega/\sqrt{\gamma_0}$.
Again, at $\omega \lesssim \gamma_{\min}$ the frequency $\omega$ in this expression should be replaced by $\gamma_{\min}$.

From this result, we find a correction to the $ 1/\omega$ noise.  This correction, 
being $\propto\omega^{-1/2}$, is due to suppression of the noise from 
TLSs that are close to resonance with the driving field $\Omega$.
The driving therefore leads to a reduction in the noise at intermediate and low frequencies.
The derivation of the expressions Eqs.~\eqref{ensdriven} and 
a similar for $\omega \lesssim \gamma_{\min}$ is a good approximation only as long as $a\ll k_BT/\hbar\sqrt{\gamma_{\min}}$. As long as this criterium holds, we find that the correction term is small relative to the first term, even if the ratio $\hbar\Omega/kT$
is small and the term $\cosh^2 (\hbar\Omega/2kT)$ approaches unity. If, however, we increase $a$ beyond this inequality, we expect that the correction term will increase until it approaches the first term in magnitude. The physics is as follows. When we increase $a$ we increase the number of fluctuators belonging to group $\MakeUppercase{\romannumeral 2}$, at the cost of group  $\MakeUppercase{\romannumeral 1}$  by increasing the width of the parabola $z^2=A^2$ in Fig.~\ref{baseintfig} until all fluctuators are resonant with the field. Thus the number of fluctuators responding to the driving field is increased and since each fluctuator within the resonant sector will have its noise spectrum shifted toward higher frequencies, the noise at low frequencies will be reduced until it approaches zero for very large fields.

\section{Decoherence of the Qubit}
\label{sec:qubit}
In this section we will describe the decoherence of the qubit due to the driven TLSs in its environment.
We will illustrate the effect from driven TLSs by use of an example with a specific, but motivated distribution of TLS parameters.
We will still assume that the qubit couples diagonally to the TLSs. The coupling was previously specified to be $H_{F-Q}=v_i\mu_z\otimes \tau_z$. Given this coupling the TLSs does only have a pure dephasing effect due to renormalization of the qubit level splitting, and direct transitions between the levels of the qubit ($T_1$ processes) cannot be induced by our two level fluctuators.

Above we have assumed that the TLSs-qubit coupling $v_i$ is not correlated with $\gamma$ and $E$. However one have to keep in mind that different distributions of $v_i$ might have significant impact upon the dephasing of the qubit. For in-depth treatment of different ensembles of fluctuators, as well as non-Gaussian noise statistics, we refer to Refs.~\onlinecite{zzbergli,schrieflshnirman,zgalperinmeso,averin}. 
Since we are in this article primarily interested in the effect of external driving, we assume that all fluctuators couple to the qubit with the same strength $v$.

In the standard Gaussian approximation, the pure dephasing time $T_2$ for long times $t$ is approximately given by $T_2^{-1}=\pi S(0)$, where $S(0)$ is the noise spectrum at zero frequency.~\cite{zzbergli}
Using this formula, together with Eq.~\eqref{totalfullsimplow} for the noise at frequencies $\omega<\gamma_{\min}$, we obtain the following expression for the dephasing time of the qubit 
\begin{align}
\frac{1}{T_2}&\approx \sqrt{8\pi}\langle v^2\rangle P_{\text{TLS}}  \!\left(  \frac{k_BT}{\gamma_{\min}}-\frac{4\hbar a}{\sqrt{\gamma_{\min}}\cosh^2\frac{\hbar\Omega}{2kT}} \right)
\label{totalfullsimplow}
\end{align}
valid for $\hbar a\gamma_{\min}\ll k T$ and $\sqrt{\gamma_0}<a<\Omega / \sqrt{\gamma_0}$.

From this expression we see that if $\hbar\Omega>k T$, then the function $\cosh^2(\hbar\Omega/2kT)\approx e^{\hbar\Omega/kT}$. We then find that the relative reduction in the noise spectrum due to driving is  $\propto\hbar a\sqrt{\gamma_{\min}}/k Te^{\hbar\Omega/kT}$. Thus the correction is exponentially suppressed at low temperatures. However we notice that if the driving frequency is reduced, i.e., if one introduce a driving field at a frequency $\hbar\Omega\leq k_BT$ much lower than the qubit frequency, the correction due to driving will become significant and the noise at zero frequency will be reduced.

\section{Discussion}
\label{discussion}
Our main result in this article, is that for Josephson junction qubits where the dominant noise source is TLSs interacting with the qubit, external driving have two main effects. The first is increased fluctuations of the TLS, contributing to increased noise at all frequencies. This effect is significant if the energy splitting of the fluctuator is small compared to temperature. The second effect is a reduction in the noise spectrum at low frequencies and at the same time increased noise at high frequencies, especially at the renormalized Rabi frequency $A'$. 

For a typical ensemble of fluctuators characterized by the distribution $P(\Delta,\Delta_0)\propto 1/\Delta_0$ we find that external driving at high frequencies (e.g., the qubit frequency) leads to reduced noise spectra at low to intermediate frequencies, which again result in an enhanced qubit dephasing time $T_2^Q$. For a typical distribution of TLSs, $P(E,\gamma)\propto 1/\gamma$, the effect is weak since the driving only reduces the noise from TLSs with energy splittings close to resonance with the driving field (group $\MakeUppercase{\romannumeral 2}$). For driving fields $\hbar\Omega>k T$, the resonant fluctuators (group $\MakeUppercase{\romannumeral 2}$) contribute only weakly to the noise also in the absence of driving. In this case, both the driven and the undriven noise spectra at low frequencies is strongly dominated by TLSs (group $\MakeUppercase{\romannumeral 1}$) with low energy splittings $E\leq k T$ and long relaxation times $T_1=1/\gamma\leq \omega^{-1}$.
However, by driving at a lower frequency $\hbar\Omega\lesssim k T$, the resonant group (group $\MakeUppercase{\romannumeral 2}$) is shifted from fluctuators close to the qubit frequency to fluctuators at lower frequencies that contribute stronger to the dephasing noise on the qubit. In this case we expect a strong reduction of the low frequency noise.

An important side effect of external driving of the TLSs that we can identify in Eq.~\eqref{corr6} is increased high frequency noise around the renormalized Rabi frequency $A'$, the driving frequency $\Omega$ and at beatings between these two frequencies. In experiments where the relaxation time $T_1^Q$ is an important limiting factor in preserving the coherence of the qubit we expect that the increased high-frequency noise due to external driving will be counterproductive and one should take measures in order to make sure that the high frequency peaks does not overlap with the qubit frequency.

\appendix
%\section{Appendixes}

\section{Averaging over the phase of the driving field}
\label{field}
The expression for the correlation function in the time domain, Eq.~\eqref{corr1}, can be written out explicitly by use of 
Eqs.~\eqref{Poperator} and~\eqref{Poperatorbar}. 
We obtain:
\begin{eqnarray}
G(t_1,t_2)&=&\left\langle F(t_2)F(t_1) \right \rangle\, , \\
F(t)&\equiv&\Delta\, \delta N(t)-\Delta_0\, \left[\delta R(t)\cos{\Omega t}-\delta I(t)\sin{\Omega t}\right]. \nonumber
\label{corra2}
\end{eqnarray}
Here we have introduced the notations $\delta N(t)=N(t)-N_{\text{ss}}$, $\delta R(t)=R(t)-R_{\text{ss}}$ and $\delta I(t)=I(t)-I_{\text{ss}}$.
This expression depends explicitly on the exact phase of the driving field, $E_{ac}$, at times $t_2$ and $t_1$.
In the following we assume that the phase of the field is random at the start of the pulse. This is typically the case if the rise time of the signal is long compared to the oscillation period $2\pi/\Omega$ of the signal. In order to average over repeated experiments with a random distribution of the phase of the field, we replace $\Omega t_1$ by $\Omega t_1+\beta$, where we assume that $\beta$ is uniformly distributed on the interval $\beta\in[0,2\pi]$.
With this assumption only products of sines and cosines contribute, while single terms vanish after averaging over the phase $\beta$.
After this averaging procedure we find that the expression for the two time correlation function Eq.~\eqref{corra2} is reduced to
\begin{eqnarray}
&&G(t,t+\tau)=(4v^2/E^2)\left\langle\Delta^2\delta N(t+\tau)\delta N(t) \right.\nonumber\\
&&\left. \quad+(\Delta_0^2/2)\left\{\delta R(t+\tau)\left[\delta R(t)\cos{\Omega\tau}+\delta I(t)\sin{\Omega\tau}\right] \right. \right.\nonumber\\
&&\left. \left. \quad\quad+\delta I(t+\tau)\left[\delta R(t)\sin{\Omega\tau}-\delta I(t)\cos{\Omega\tau}\right]\right\}\right\rangle
\label{corr3}
\end{eqnarray}
where the cross terms of Eq.~\eqref{corra2} proportional to $\Delta\Delta_0$ have canceled due to averaging over repeated experiments with random distribution of the initial phase of the driving field.

\section{Coordinate transformation and averaging over initial conditions} 
\label{App:init}
The Bloch-Redfield equations are equations of motion for the average of an ensemble of TLSs, where the individual members of the ensemble differ by details of the environment. By use of the Bloch-Redfield equations we avoid dealing with these details that we do not have control over. In place we get an equation of motion for the mean density matrix. While this method greatly simplifies the dynamics, since we are no longer required to keep track of fine details of the environment, the cost is loss of information about the time evolution of individual systems. Therefore the Bloch-Redfield equations cannot be used to calculate two time correlation functions in a straightforward way.~\cite{laxrmp,laxflucdrivstate} The procedure we use in order to evaluate the two-time correlation function is the following.
In general the two-time correlation function can be expressed  as
$$ 
\langle A(t_1)A(t_2)\rangle=\sum\limits_{j,k}
a_k(t_2)\zeta(a_k(t_2)|_{a_j(t_1)})a_j(t_1)\xi(a_j(t_1)).
$$ 
Here $A(t)$ and $a(t)$ is an observable and a particular realization of this observable, respectively, $\zeta(a_k(t_2)|_{a_j(t_1)})$ is the conditional probability distribution for observing the value $a_k(t_2)$ at time $t_2$ conditioned upon that the value $a_j(t_1)$ was observed at $t_1$. And the $\xi(a_j(t_1))$ is simply the probability distribution for observing the value $a_j(t_1)$ at time $t_1$.
We might then realize that $\sum\limits_ka_k(t_2)\zeta(a_k(t_2)|_{a_j(t_1)})=\langle A(t_2)\rangle |_{a_j(t_1)}$ is simply the solution of the Bloch-Redfield equation given the initial value $a_j(t_1)$.
The two-time correlation function thus reduces to
\begin{align}
\langle A(t_1)A(t_2)\rangle&=\sum\limits_j\langle A(t_2)\rangle |_{a_j(t_1)}a_j(t_1)\xi(a_j(t_1)).
\label{generalcorr}
\end{align}
 If we now move to our specific problem of a driven TLS in a dissipative environment, the two time correlation function we require, Eq.~\eqref{corr1}, does only contain terms with the deviation of the observable quantity from its steady state value. From the explicit solutions of the Bloch-Redfield equations, Eq.~\eqref{transient} and Eq.~\eqref{steadystate},  we see that the time evolution of the deviation from the steady state is translation invariant, linear and homogeneous.
Therefore, we can write 
\begin{widetext}
\begin{eqnarray}
&&\langle q(t_2)-\bar{q}(t_2)\rangle |_{q_j(t_1)-\bar{q}(t_1)}= 
(2d/E)\big\{[\Delta\lambda_{11}(\tau)-\Delta_0(\lambda_{21}(\tau)\cos{\Omega\tau}-\lambda_{31}(\tau)\sin{\Omega\tau})]
(N_j(t_1)-N_{\text{ss}}) + [\Delta\lambda_{12}(\tau) \\
&&-\Delta_0(\lambda_{22}(\tau)\cos{\Omega\tau}
-\lambda_{32}(\tau)\sin{\Omega\tau})]
(R_j(t_1)-R_{\text{ss}})
+ [\Delta\lambda_{13}(\tau)-\Delta_0(\lambda_{23}(\tau)\cos{\Omega\tau}-\lambda_{33}(\tau)\sin{\Omega\tau})]
(I_j(t_1)-I_{\text{ss}})\big\}. \nonumber
\end{eqnarray}
\end{widetext}
Here $\tau=t_2-t_1$ and $\lambda_{\alpha\beta}(\tau)$ are the elements of $\Lambda(\tau)$ given by Eq~\eqref{transient}.
This expression can be inserted directly into our general formula given by Eq.~\eqref{generalcorr}.
Unfortunately we still require the distribution function $\xi(N_j(t_1))$, and similarly, the distribution of $I$ and $R$. The distribution functions cannot be extracted from the Bloch-Redfield equations.~\cite{laxflucdrivstate}. To make up for our lack of information, we make the following approximation.
Assuming that a measurement of the TLS in the eigenbasis of $H_0$, will give either the outcome $N_{j=0}=0$ or $N_{j=1}=1$ with the mean value $N_{\text{ss}}$, we find that 
\begin{align}
\sum\limits_jN_j\xi(N_j)&=N_0\xi(N_0)+N_1\xi(N_1)=N_{\text{ss}},\nonumber\\
\sum\limits_jN_j^2\xi(N_j)&=N_0^2\xi(N_0)+N_1^2\xi(N_1)=N_{\text{ss}}.
\label{summation}
\end{align}
In addition, with this choice of initial values, the initial values of the off-diagonal density matrix elements is always zero. 
We note that this choice of initial values only make sense if the steady state density matrix lie on the axis between the points $N_{j=0}=0$ or $N_{j=1}=1$. In order to apply the method, we are therefore required to transform to the coordinate system where the steady state value of $\rho$ lie on the z-axis of the Bloch sphere.

\subsection{Coordinate transformation}
The summation procedure derived above, Eq.~\eqref{summation}, works nicely as long as the assumptions behind the derivation of the Bloch-Redfield equations (the Born-Markov approximations) are fulfilled~\cite{slichter,redfield}, as well as the time translation invariance. However, the summation procedure can only be applied if the steady state of the density matrix lies along the $z$-axis of the relevant measurement operator.
With driving, we see that the density matrix will in general be driven away from the $z$-axis, such that a summation over the eigenstates $|+\rangle$ and $|-\rangle$ of the $H_0$ operator given by Eq.~\eqref{Ham}, can not possibly give the true average density matrix in steady state.
However, we can do proper averaging by transforming to a new coordinate system where $N'_{++}$ denotes the occupation along the $z'$-axis in this new choice of coordinates. After this transformation the Bloch vector of the steady state density matrix $N_{\text{ss}}$ is a point on this axis. By this choise of axes the density matrix is diagonal. 
The coordinate transformation is given as 
\begin{eqnarray}
&&(N-1/2)=(N'-1/2) \cos\theta, 
\nonumber \\
&& R=N'\sin\theta \sin\phi,  \ I=N'\sin\theta \cos\phi .
\label{transf}
\end{eqnarray}
We note that $R'_{\text{ss}}=0$ and $I'_{\text{ss}}=0$ since we have defined $N'_{\text{ss}}$ to lie on the $z'$-axis in the new coordinate system.
When we insert the explicit steady state expressions into Eq.~\eqref{ctrans}, the dependence on the equilibrium value $N_{eq}$ vanishes, and the expressions reduce to
\begin{equation} \label{angpar}
\tan \theta= \frac{a^2\gamma(z^2+\gamma2)}{4(\gamma^2+z^2+2a^2\gamma)^2}, \quad \tan \phi=\frac{z}{\gamma}\, .
\end{equation}

Inserted into our expression for the correlation function Eq.~\eqref{corr3} we obtain the following formula:
\begin{widetext}
\begin{eqnarray}
&G(t_1,t_2)&= (2v^2/E^2)\big\langle 2 \Delta^2\delta N(t_1)
[\lambda_{11}(\tau)\delta N(t_1)+\lambda_{12}(\tau)\delta R(t_1)+\lambda_{13}(\tau)\delta I(t_1)]
+\Delta_0^2\big\{[\lambda_{21}(\tau)\delta N(t_1)+\lambda_{22}(\tau)\delta R(t_1)
\nonumber\\
& &\quad+\lambda_{23}(\tau)\delta I(t_1)]
[\delta R(t_1)\cos{\Omega\tau}+\delta I(t_1)\sin{\Omega\tau}]
+[\lambda_{31}(\tau)\delta N(t_1)+\lambda_{32}(\tau)\delta R(t_1)+\lambda_{33}(\tau)\delta I(t_1)]
\nonumber\\
&&\quad \quad\times
[\delta R(t_1)\cos{\Omega\tau}+\delta I(t_1)\sin{\Omega\tau}]\big\}\big\rangle .
\label{corrformula}
\end{eqnarray}
We can now by use of Eq.~\eqref{transf} move to the frame where the density matrix lie along the $z$-axis.
By use of the summation formulas given by Eq.~\eqref{generalcorr} and Eq.~\eqref{summation}, and inserting the explicit expressions for the elements of $\Lambda(\tau)$, given by Eq.~\eqref{transient} we obtain the following expression for the correlation function:
\begin{eqnarray}
&&G(\tau,0)=\left(2v/EA'\right)^2 N'_{\text{ss}}(1-N'_{\text{ss}})\big\{\Delta^2[(A^2\cos{A'\tau}+z^2)\cos^2\theta-zA(1-\cos{A'\tau})\cos\theta\sin\theta\sin\phi\nonumber\\ 
&&\quad +AA'\sin{A'\tau}\cos\theta\sin\theta\cos\phi]+
\frac{\Delta_0^2}{2}\big[A'^2\sin^2\theta\cos\phi\cos{A'\tau}\cos{\Omega\tau}(\cos\phi-\sin\phi)+AA'\cos\theta\sin\theta\sin{A'\tau}\nonumber\\ 
&&\quad\quad \times (\sin\phi\sin{\Omega\tau}-\cos\phi\cos{\Omega\tau})
+\frac{z}{A'}\sin^2\theta\sin{A'\tau}\sin{\Omega\tau}
+\frac{A^2}{A'^2}\sin^2\theta\sin\phi(\cos\phi\sin{\Omega\tau}-\sin\phi\cos{\Omega\tau})\nonumber\\ 
&&\quad\quad \quad +z^2\sin^2\theta\sin\phi\sin{A'\tau}(\cos\phi\sin{\Omega\tau}-\sin\phi\cos{\Omega\tau})
-zA\cos\theta\sin\theta(\cos\phi\sin{\Omega\tau}+\sin\phi\cos{\Omega\tau})\nonumber\\ 
&&\quad\quad \quad \quad +zA\cos\theta\sin\theta\cos{A'\tau}(\cos\phi\sin{\Omega\tau}-\sin\phi\cos{\Omega\tau})\big]\big\}.
\label{corrfinal}
\end{eqnarray}
Here 
\begin{eqnarray}
&&N'_{\text{ss}}(1-N'_{\text{ss}})=N_{\text{ss}}(1-N_{\text{ss}})+\left(N_{\text{ss}}-\frac{1}{2}\right)^2\sin^2 \theta -\frac{R_{\text{ss}}^2}{4}\sin^2 \theta \sin^2 \phi -\frac{I_{\text{ss}}^2}{4}\sin^2 \theta \cos^2 \phi \nonumber\\
&&-\left(N_{\text{ss}}-\frac{1}{2}\right) R_{\text{ss}}\cos \theta \sin \theta \sin \phi -\left(N_{\text{ss}}-\frac{1}{2}\right)I_{\text{ss}}\cos(\theta)\sin \theta \cos \phi -\frac{I_{\text{ss}}R_{\text{ss}}}{2}\sin^2 \theta \cos \phi \sin \phi .
\label{nss}
\end{eqnarray}
Eq.~\eqref{corrfinal} is our final expression for the two time correlation function for a single fluctuator. We note that the correlation function has become fully translation invariant after the averaging procedure. The Fourier spectrum is computed in Appendix~\ref{fullspectrum}.

\section{Spectral density $S(\omega)$ at arbitrary frequency}
\label{fullspectrum}
In this appendix we give the full spectral density from a single TLS.
The full spectrum is given by the fourier transform with respect to $\tau=t_2-t_1$ of Eq.~\eqref{corrfinal}.
Thus we calculate $S(\omega)=\sqrt{2/\pi}\int\limits_{-\infty}^{\infty}e^{i\omega\tau} G(|\tau|,0) d\tau$.
Carrying out the transform we obtain the result
\begin{eqnarray} 
&S(\omega)&=8\sqrt{\frac{2}{\pi}}\left(\frac{v}{EA'}\right)^2N'_{\text{ss}}(1-N'_{\text{ss}})\Big(\Delta^2\Big\{A^2\cos^2{\theta}+zA\sin{\theta}\cos{\theta}\sin{\phi}[L(\omega+A')+L(\omega-A')]
\nonumber\\ 
&&\quad\quad +AA'\sin{\theta}\cos{\theta}\cos{\phi}\Big[\frac{A'+\omega}{\gamma}L(\omega+A')+\frac{A'-\omega}{\gamma}L(\omega-A')\Big]+(z^2\cos^2{\theta}-zA\sin{\theta}\cos{\theta}\sin{\phi})L(\omega)\Big\}\nonumber\\ 
&&\quad+\frac{\Delta_0^2}{4}\sin{\theta}\Big\{b_{1-}[L(\omega+A'+\Omega)+L(\omega-A'-\Omega)]+b_{1+}[L(\omega+A'-\Omega)+L(\omega-A'+\Omega)]\nonumber\\ 
&&\quad\quad +b_{2+}\Big[\frac{A'+\Omega+\omega}{\gamma}L(\omega+A'+\Omega)+\frac{A'+\Omega-\omega}{\gamma}L(\omega-A'-\Omega)\Big]\nonumber\\ 
&&\quad\quad +b_{2-}\Big[\frac{A'-\Omega+\omega}{\gamma}L(\omega+A'-\Omega)+\frac{-A'+\Omega+\omega}{\gamma}L(\omega-A'+\Omega)\Big]\nonumber\\
&&\quad\quad +2b_3\Big\{\sin\phi [L(\omega+\Omega)+L(\omega-\Omega)]+\cos\phi \Big[\frac{\Omega+\omega}{\gamma}L(\omega+\Omega)+\frac{\Omega-\omega}{\gamma}L(\omega-\Omega)\Big]\Big\}\Big\}\Big),\label{corr6}\\
& b_{1\pm}&=A'^2\sin{\theta}\cos{\phi}(\cos{\phi}-\sin{\phi})\pm AA'\cos{\theta}(1+\cos{\phi})+z\sin{\theta}+z^2\sin{\theta}\sin{\phi}+Az\cos{\theta})(\cos{\phi}+\sin{\phi}),\nonumber\\
& b_{2\pm}&=-AA'\cos\theta\cos\phi\pm z^2\sin{\theta}\sin{\phi}\pm Az\cos{\theta})(\cos{\phi}+\sin{\phi}),\nonumber\\ 
& b_3&=A^2\sin\theta\sin\phi-Az\cos\theta.\nonumber\\
\end{eqnarray}
\end{widetext}

From Eq.~\eqref{corr6}, we see that without driving (and the same for off-resonant driving), we have a single peak 
$$S(\omega)\propto\frac{z^2\cos^2{\theta}-zA\sin{\theta}\cos{\theta}\cos{\phi}}{A^2+z^2}\frac{\gamma}{\gamma^2+\omega^2}$$ that is reduced as the driving increases (i.e., when the Rabi frequency $A$ approaches $z$).
When the driving is strong, the intensity is shifted from the single peak at $\omega=0$, to a large number of peaks. Most prominent are the peaks at the renormalized Rabi frequency, $\omega=A'$, but there are also peaks at the driving frequency, and at sums and differences between $A'$ and $\Omega$.
The noise at $\omega\approx\Omega$ has not been discussed in this article, but might be important for qubit operation if the driving field is the manipulating pulses used to control the qubit. This noise is then close to resonance with the qubit.

\section{Derivation of the ensemble averaged noise for strong driving}
\label{strongdriving}

\begin{figure}[htb]
  \includegraphics[width=0.4\textwidth]{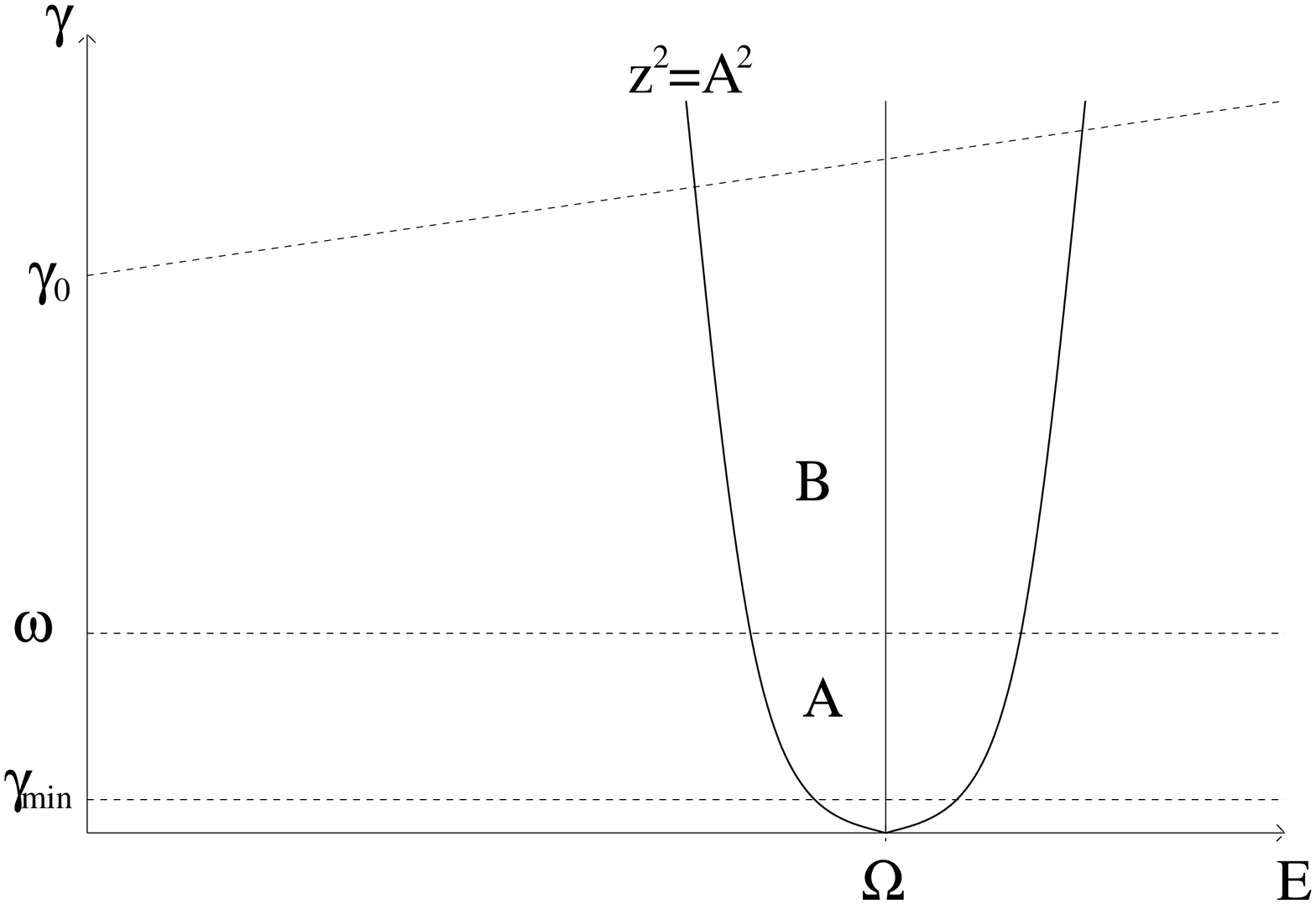}
 \caption{The domains of integration}
  \label{intfigapp}
\end{figure}

In this section we will derive the ensemble averaged spectrum of the low frequency noise  induced by strongly driven TLSs (i.e., $a^2 \gg \gamma_0$) for the distribution of TLS parameters given by Eq.~\eqref{dist}.
In order to evaluate the integral over the TLS parameters we make the following approximation. The full parameter domain $\gamma\otimes E\in[\gamma_{\min},\gamma_0]\otimes[0,\infty]$ is split in two sectors. In the first sector the TLSs are in resonance with the driving field defined by the criteria $a^2\gamma>z^2$ (group $\MakeUppercase{\romannumeral 2}$). 
The second one contains the TLSs that are out of resonance (group $\MakeUppercase{\romannumeral 1}$, defined by $a^2\gamma<z^2$). The domains of integration is specified in Fig.~\ref{intfigapp}. For the fluctuators belonging to the resonant sector we use the expression for the noise spectrum in resonant strong driving given by Eq.~\eqref{strong2}, while in the off resonant sector we use the expression for fluctuators out of resonance given by \eqref{zexpy}. 
Our strategy to evaluate the noise spectra is to first compute the ensemble averaged noise spectrum in the absence of driving over the full domain, then subtract the contribution to the undriven spectrum from fluctuators that lie in the resonant sector, and finally we add the contribution from the resonant fluctuators in group $\MakeUppercase{\romannumeral 2}$.
Therefore, the noise spectrum can be represented as
\begin{equation}
\bar{S}_{a^2>\gamma}(\omega)=
\bar{S}_{0}(\omega)-\bar{S}_{0}^{(res)}(\omega)+\bar{S}_{a^2>\gamma}^{(res)}(\omega).
\label{intsum}
\end{equation}
Here the superscript denote the resonant sector.

We start by evaluating the spectrum in the absence of driving, by use of Eqs.\eqref{zexpy} and \eqref{dist} we get
\begin{equation}
\bar{S}_{0}(\omega)
=\sqrt{\frac{8}{\pi}}\langle v^2\rangle \int\limits_0^{\infty}\! \int\limits_{\gamma_{\min}}^{\gamma_0} \! \! \frac{P_{\text{TLS}}\sqrt{1-\frac{\gamma}{\gamma_0}}}{(\gamma^2+\omega^2)\cosh^2{\frac{E}{2kT}}}dEd\gamma.
\label{freeens}
\end{equation}
For $\gamma_{\min}\ll \omega \ll \gamma_0$ the integral over $\gamma$ can be estimated as
\begin{equation}
\int\limits_{\gamma_{\min}}^{\omega}\frac{d \gamma}{\omega^2}+\int\limits_{\omega}^{\gamma_0}\frac{1-\frac{\gamma}{2\gamma_0}}{\gamma^2}d\gamma
\approx \frac{2}{\omega}-\frac{\gamma_{\min}}{\omega^2}-\frac{1}{\gamma_0}-\frac{1}{2\gamma_0}\ln{\frac{\gamma_0}{\omega}}, \nonumber
\end{equation}
while the energy integral is $\int\limits_{0}^{\infty} dE \, \cosh^{-2} \!\left(\frac{E}{2kT}\right)=2kT$.
Therefore
\begin{equation}
\bar{S}_{0}(\omega)=\mathcal{A}
\Big(\frac{2}{\omega}-\frac{\gamma_{\min}}{\omega^2}
-\frac{1}{\gamma_0}-\frac{1}{2\gamma_0}\ln{\frac{\gamma_0}{\omega}}\Big). \nonumber
\label{ens0}
\end{equation}
where $\mathcal{A}=\sqrt{8/\pi}\langle v^2\rangle P_{\text{TLS}}k T$.
For very low frequencies $\omega<\gamma_{\min}$ we get
\begin{equation}
\bar{S}_{a=0}(\omega)\approx \mathcal{A}\Big(\frac{1}{\gamma_{\min}}
-\frac{1}{\gamma_0}-\frac{1}{2\gamma_0}\ln{\frac{\gamma_0}{\gamma_{\min}}}\Big).
\label{ens0low}
\end{equation}

Next we proceed to calculate the contribution from the resonant sector in the absence of driving.
The integral we need to evaluate is similar to that of Eq.~\eqref{freeens}, but the integration should be performed over the TLSs from group $\MakeUppercase{\romannumeral 2}$.
The integration domain is restricted by the parabola $z^2\leq a^2\gamma$.

Assuming first that $\gamma_{\min}<\omega<\gamma_0$, we evaluate the integral by use of asymptotic expressions for the integrand in the different regions of Fig. \ref{intfigapp}. More precisely we use $\frac{1}{\gamma^2+\omega^2}\approx\frac{1}{\omega^2}$ in sector $A$ of Fig.\ref{intfigapp}, and $\frac{1}{\gamma^2+\omega^2}\approx\frac{1}{\gamma^2}$ in sector $B$. 
Furthermore we make the approximation $\cosh^2{\frac{E}{2k T}}\approx \cosh^2{\frac{\hbar\Omega}{2k_BT}}=\text{const}$ for all fluctuators inside the resonant sector. If we write $E=\hbar(\Omega-z)$, we find that since $z^2\leq a^2\gamma_0$ inside the resonant sector, the approximation is good as long as $\frac{\hbar z}{k_BT}\leq\frac{\hbar a\sqrt{\gamma_0}}{k T}\ll 1$. However, the major contribution to the integral comes from $\gamma\lesssim\omega$, such that we can narrow our inequality to  $\frac{\hbar z}{k T}\leq\frac{\hbar a\sqrt{\omega}}{k T}\ll 1$.
By use of the approximations described above, we find that the total contribution from the resonant sector in the absence of driving, in the frequency interval $\gamma_{\min}<\omega<\gamma_0$, is 
\begin{equation}
\bar{S}_{0}^{(res)}
=\mathcal{A}
\frac{2\hbar a}{kT \cosh^2{\frac{\hbar\Omega}{2kT}}}\! \!
\left(\frac{8}{3\sqrt{\omega}}-\frac{3}{\sqrt{\gamma_0}}-\frac{2\gamma_{\min}^{\frac{3}{2}}}{3\omega^2}+\frac{\sqrt{\omega}}{\gamma_0}\right).
\label{res0}
\end{equation}
For $\omega<\gamma_{\min}$, the calculations are similar, but somewhat simpler.
The result is 
\begin{equation}
\bar{S}_{0}^{(res)}=\mathcal{A}\frac{2\hbar a}{kT \cosh^2{\frac{\hbar\Omega}{2kT}}}
\left(\frac{2}{\sqrt{\gamma_{\min}}}-\frac{3}{\sqrt{\gamma_0}}+\frac{\sqrt{\gamma_{\min}}}{\gamma_0}\right).
\label{res0low}
\end{equation}

Finally we proceed to the resonant sector in strong external driving, $a^2>\gamma$:
\begin{equation}
\bar{S}_{a^2>\gamma}^{(res)}(\omega)
=\frac{5\mathcal{A}}{2a^2 k T}\iint\limits_{group\; \MakeUppercase{\romannumeral 2}}\frac{\sqrt{1-\frac{\gamma}{\gamma_0}}}{\gamma}\, d\gamma dE,
\label{ensinr}
\end{equation}
 where we have used the expression for the resonant spectrum in strong driving given by Eq.~\eqref{strong2}.
This integral is evaluated similarly to the corresponding integral for the same region in the absence of driving.
After using asymptotic expressions in the different regions of Fig.~\ref{intfigapp}, we find 
\begin{equation}
\bar{S}_{a^2>\gamma}^{(res)}(\omega)
\approx \frac{25\mathcal{A}\hbar\sqrt{\gamma_0}}{16a k T}.
\label{agammaint3}
\end{equation}
for the total contribution from the resonant sector in external field.
This expression is valid for $\omega<\gamma_0$.

We have now computed the three contributions to the total ensemble averaged noise spectrum given by Eq.~\eqref{intsum}. For $\gamma_{\min}<\omega<\gamma_0$ the full spectrum is given by
\begin{eqnarray}
\bar{S}_{a^2>\gamma}(\omega)
&\approx& \mathcal{A} 
\Big[\Big(\frac{2}{\omega}-\frac{\gamma_{\min}}{\omega^2}
-\frac{1}{\gamma_0}-\frac{1}{2\gamma_0}\ln{\frac{\gamma_0}{\omega}}\Big)
\nonumber\\ 
&&  +\frac{25\hbar\sqrt{\gamma_0}}{4akT}
- \frac{2\hbar a}{kT\cosh^2{\frac{\hbar\Omega}{2kT}}}\left(\frac{8}{3\sqrt{\omega}} \right.
\nonumber\\ 
&& \left .\quad 
-\frac{3}{\sqrt{\gamma_0}}-\frac{2\gamma_{\min}^{3/2}}{3\omega^2}+\frac{\sqrt{\omega}}{\gamma_0}\right)\Big].
\label{totalfullcorr}
\end{eqnarray}
Within the limits of the inequality used for the evaluation of the integrals, $\hbar a\sqrt{\omega}\ll k_BT$, and by using that $a>\sqrt{\omega}$, we see that the second term origining from the driven resonant sector is negligible compared to the undriven $1/\omega$ term.
Thus we are left with 
\begin{eqnarray}
&&\bar{S}_{a^2>\gamma}(\omega) \approx \mathcal{A}
\Big[\Big(\frac{2}{\omega}-\frac{\gamma_{\min}}{\omega^2}-\frac{1}{\gamma_0}(1+\ln{\sqrt{\frac{\gamma_0}{\omega}}}\Big)
\nonumber\\ 
&& -\frac{2\hbar a}{kT \! \cosh^2 \! \! \frac{\hbar\Omega}{2kT}}\! \! \left(\! \! \frac{8}{3\sqrt{\omega}}-\frac{3}{\sqrt{\gamma_0}}-\frac{2\gamma_{\min}^{3/2}}{3\omega^2}+\frac{\sqrt{\omega}}{\gamma_0} \! \right) \! \!\Bigg]\! .
\label{totalfullsimp}
\end{eqnarray}

For frequencies $\omega<\gamma_{\min}$ the contributions to the ensemble averaged spectrum is given by Eqs.\eqref{ens0low},\eqref{res0low} and \eqref{agammaint3}, giving:
\begin{eqnarray}
&& \bar{S}_{a^2>\gamma}(\omega)
\approx \mathcal{A}\Big[\Big(\frac{1}{\gamma_{\min}}-\frac{1}{\gamma_0}
-\frac{1}{2\gamma_0}\ln{\frac{\gamma_0}{\gamma_{\min}}}\Big)
\nonumber\\ 
&& -\frac{2\hbar a}{kT \cosh^2 \! \frac{\hbar\Omega}{2kT}} \!\left(\! \frac{2}{\sqrt{\gamma_{\min}}}-\frac{3}{\sqrt{\gamma_0}}+\frac{\sqrt{\gamma_{\min}}}{\gamma_0}\! \right) \! \!\Big] \! .
\label{totalfullsimplow}
\end{eqnarray}

\end{document}